\documentstyle[12pt]{article}
\begin{document}
\begin{center}
{\LARGE {Charging Symmetries and Linearizing Potentials\\
\vskip 0.2cm
for Einstein-Maxwell Dilaton-Axion Theory
}}
\end{center}
\vskip 1.5cm
\begin{center}
{\bf \large {Alfredo Herrera-Aguilar}}
\end{center}
\begin{center}
Joint Institute for Nuclear Research,\\
Dubna, Moscow Region 141980, RUSSIA.\\
e-mail: alfa@cv.jinr.dubna.su
\end{center}
\vskip 0.1cm
\begin{center}
and
\end{center}
\vskip 0.1cm
\begin{center}
{\bf \large {Oleg Kechkin}}
\end{center}
\begin{center}
Institute of Nuclear Physics,\\
Moscow State University, \\
Moscow 119899, RUSSIA, \\
e-mail: kechkin@monet.npi.msu.su
\end{center}
\vskip 1.5cm
\begin{abstract}
We derive a set of complex potentials which linearize the action of
charging symmetries of the stationary Einstein-Maxwell
dilaton-axion theory.
\end{abstract}
\newpage
\section{Introduction}
Superstring theory provides a correct quantum description of gravity
coupled to matter fields. In the low energy limit superstring
theory leads to some modifications of General Relativity. These
$string$ $gravity$ $models$ preserve long-distance behaviour of
the mysterious quantum gravity and in special (BPS-saturated) cases
exactly reproduce it \cite {kir}.

Einstein-Maxwell theory with dilaton and axion fields (EMDA) is
one of the simplest string gravity models. It is described by the action
\begin{eqnarray}
S = \int d^4 x |g|^{\frac {1}{2}} \left \{ -R + 2(\partial \phi)^2
+ \frac{1}{2}e^{2\phi}(\partial \kappa)^2
- e^{-2\phi}F^2 - \kappa F\tilde F \right \},
\nonumber
\end{eqnarray}
where $F_{\mu \nu} = \partial _{\mu}A_{\nu} - \partial _{\nu}A_{\mu}$
is the Maxwell strength and $\tilde F^{\mu \nu} = \frac {1}{2}
E^{\mu \nu \lambda \sigma}F_{\lambda \sigma}$. Formally EMDA can be
considered as an extension of the Einstein-Maxwell
(EM) theory to the case of non-trivial scalar dilaton field
$\phi$ and pseudoscalar axion field $\kappa$.

EMDA arises in the framework
of the non-critical heterotic string theory (D=4, one vector field).
It can be also obtained as the corresponding truncation of the
critical heterotic string theory (D=10, 16 vector fields) reduced to
four dimensions. The EMDA solution spectrum
and symmetry structure were under extensive
investigation during last several years (see \cite {youm} for
a review).

In the stationary case the symmetry group of this theory consists
of gauge and non-gauge parts. The first part is trivial, whereas the second
one
acts in the charge space of asymptotically flat field configurations.
We name non-gauge transformations
as $charging$ $symmetries$ because they generate charged solutions from
neutral ones.

In this letter we derive a representation of the stationary EMDA
using complex potentials which transform linearly under the action
of the charging symmetries. The found potentials provide an adequate
description of the EMDA field configurations possessing the flatness
property at spatial infinity.
\section{Charging symmetries}
In this section we review the matrix Ernst potential formulation \cite {gk}
and list charging symmetries of the stationary EMDA \cite {ky}. At the end
of the section the main problem of the letter is formulated.

Our notations are the following. For the D=4 line element we use the
decomposition
\begin{eqnarray}
ds^2 = f(dt - \omega _i dx^i)^2 - f^{-1} h_{ij}dx^idx^j;
\nonumber
\end{eqnarray}
thus $f, \omega_i$ and $h_{ij}$ become scalar, vector and symmetric
tensor 3-fields in the stationary case. Next, we introduce the magnetic
potential $u$ on shell
\begin{eqnarray}
\nabla u = \sqrt 2 \left \{ fe^{-2\phi}\left ( \nabla \times \vec A +
\nabla A_0 \times \vec \omega \right ) + \kappa \nabla A_0 \right \}.
\nonumber
\end{eqnarray}
The electric potential $v$ is  $\sqrt 2 A_0$. Finally, the rotational
potential $\chi$ can be defined exactly as in the EM theory
\cite {iw}:
\begin{eqnarray}
\nabla \chi &=& u\nabla v - v\nabla u - f^2 \nabla \times \vec \omega.
\nonumber
\end{eqnarray}
A K\"ahler formulation of the theory is based on the use of three complex
functions
\begin{eqnarray}
{\cal Z} &=& \kappa + ie^{-2\phi},
\nonumber \\
{\cal F} &=& u - {\cal Z}v,
\nonumber \\
{\cal E} &=& if - \chi + v{\cal F}.
\nonumber
\end{eqnarray}
which generalize Ernst potentials of the EM theory \cite {ernst1}.
They can be combined into the matrix
\begin{eqnarray}
E = \left (\begin{array}{crc}
{\cal E}&\quad &{\cal F} \\
{\cal F}&\quad & -{\cal Z}\\
\end{array}\right ),
\nonumber
\end{eqnarray}
which we call $matrix$ $Ernst$ $potential$ (MEP) \cite {gk}
in view of the close analogy between EMDA in terms of MEP and Einstein
gravity using the Ernst potential formulation \cite {ernst2}.
Actually, the effective
EMDA action in the stationary case reads:
\begin{eqnarray}
S = \int d^3 x h^{\frac {1}{2}} \left \{ -^3R +
2{\rm Tr} \left ( J^EJ^{\bar E} \right ) \right \}.
\nonumber
\end{eqnarray}
where $J^E=\nabla E (E-\bar E)^{-1}$; it concides with Einstein's one
if matrix $E$ is replaced by a function.

Using the MEP representation the EMDA symmetries
adopt the ``matrix valued
SL(2,R) form'':
\begin{eqnarray}
E \rightarrow S^T(E^{-1}+L)^{-1}S +R,
\end{eqnarray}
where ${\rm det}S\neq 0$ and matrices $L$ and $R$ are symmetric
\cite {gk}. Eq. (1) describes the full symmetry group, the so-called
U-duality. In this work we shall consider the charging subgroup.

Matrices $S$, $L$ and $R$ of this subgroup satisfy the relation
\begin{eqnarray}
\sigma_3+iR=S^T(\sigma_3+iL)^{-1}S
\end{eqnarray}
It provides the conservation of the vacuum solution $E_{vac}=i\sigma_3$
(all the 4D matter fields are trivial and the metric corresponds to the
Minkowski space-time).

In \cite {ky} it was shown how to solve Eq. (2). The solution
consists of two commuting parts. The first part is
\begin{eqnarray}
S_{U(1)}=(\cos \lambda ^0)^{-1}\sigma _0,\qquad
R_{U(1)}=-L_{U(1)}=\sigma _3\tan \lambda ^0;
\end{eqnarray}
The second one is
\begin{eqnarray}
S_{SU(1,1)} &=& \Delta ^{-1}[f_1\sigma _0+\lambda ^1f_2\sigma _1],
\nonumber \\
R_{SU(1,1)} &=& f_2\Delta ^{-1}[(\lambda ^3f_1+\lambda ^1\lambda ^2f_2)\sigma _0+
(\lambda ^2f_1+\lambda ^1\lambda ^3f_2)\sigma _1],
\nonumber \\
L_{SU(1,1)} &=& f_2\Delta ^{-1}[(-\lambda ^3f_1+\lambda ^1\lambda ^2f_2)\sigma _0+
(\lambda ^2f_1-\lambda ^1\lambda ^3f_2)\sigma _1],
\end{eqnarray}
where $\Delta =f_1^2 -(\lambda^1)^2f_2^2$, $\lambda ^{\mu}$  are group
parameters and
\begin{eqnarray}
2f_1 &=& (1+\sigma)\cosh \lambda +(1-\sigma)\cos \lambda,
\nonumber \\
2\lambda f_2 &=& (1+\sigma)\sinh \lambda +(1-\sigma)\sin \lambda.
\end{eqnarray}
Here we have introduced the parameter
$\lambda =\sqrt{|(\lambda^1)^2+(\lambda^2)^2-(\lambda^3)^2|}$
and put $\sigma ={\rm sign} [(\lambda^1)^2+(\lambda^2)^2-(\lambda^3)^2]$.

Let us consider field configurations near the vacuum point. Then
$E_{as}=i(\sigma _3-2\hat {\cal M}/r)$,
where $1/r$ is the infinitesimal parameter (in what follows $r$ tends to
spatial infinity), and
\begin{eqnarray}
{\hat {\cal M}}=\left (\begin{array}{crc}
{\cal M}&{\cal Q}\\
{\cal Q}&-{\cal D}\\
\end{array}\right )
\nonumber
\end{eqnarray}
is a constant matrix. Its components are ${\cal M}=M+iN$ (M is the
Arnowitt-Deser-Misner mass, N denotes the Newmann-Unti-Tamburino parameter),
${\cal D}=D+iA$ (a combination of dilaton and axion charges) and the
electromagnetic charge ${\cal Q}=Q_e+iQ_m$. A remarkable fact
is that ${\hat {\cal M}}$ components transform linearly
under the action of the charging symmetries. Actually, in
\cite {ky} it was shown that
\begin{eqnarray}
{\hat {\cal M}}\rightarrow T^T{\hat {\cal M}}T
\end{eqnarray}
where
$T=(1+iL\sigma_3)^{-1}S$.
The operator $T$ is the commuting product of the operators
\begin{eqnarray}
T_{U(1)} = e^{i\lambda^0}\sigma_0 \qquad {\rm and} \qquad
T_{SU(1,1)} = f_1\sigma_0+f_2(\lambda^1\sigma_1-i\lambda^2\sigma_2+
i\lambda^3\sigma_3),
\end{eqnarray}
i.e.
\begin{eqnarray}
T=T_{U(1)}T_{SU(1,1)}=T_{U(1)}T_{SU(1,1)};
\end{eqnarray}
it satisfies the relation
\begin{eqnarray}
T^+\sigma_3T=\sigma_3.
\end{eqnarray}

Now we are ready to formulate the problem of this work. One can see that
both sets of variables $({\cal E}, {\cal Z}, {\cal F})$ and $({\cal M},
{\cal D}, {\cal Q})$ realize some representations of the charging
subgroup.
Variables of the first set transform in a highly complicated way, whereas
the second set transforms linearly. It is natural to suppose that
other potentials which transform like charges exist. The derivation
of these potentials is presented in the next section.
\section{Linearizing potentials}
Let us denote the complex potentials which linearize the action of charging
symmetries as $w_1$, $w_2$ and $w_3$ and combine them into the
symmetric matrix
\begin{eqnarray}
\hat {\cal W}=\left (\begin{array}{crc}
w_1&w_3\\
w_3&w_2\\
\end{array}\right ),
\nonumber
\end{eqnarray}
This matrix transforms as
\begin{eqnarray}
\hat {\cal W}\rightarrow T^T\hat {\cal W}T
\end{eqnarray}
under the charging subgroup.

Our plan is the following: to establish the relation between EMDA potentials
and the unknown variables
\begin{eqnarray}
w_i=w_i({\cal E}, {\cal Z}, {\cal F}),
\end{eqnarray}
we calculate the infinitesimal generators of the charging symmetries in
both representations and identify these generators using their commutation
relations. This gives a set of differential equations which defines the
relations (11).

A great simplification of the solution process can be achieved if one
works with appropriate variables. After some algebraic manipulations we lead to
the sets $({\cal E}_+, {\cal E}_-, {\cal F})$, where
\begin{eqnarray}
{\cal E}_{\pm}=\frac {1}{2}({\cal E}\pm {\cal Z}),
\end{eqnarray}
and $(r, \theta, \varphi)$, where
\begin{eqnarray}
\frac {1}{2}(w_1+w_2)&=&r\sinh \theta \cos \varphi,
\nonumber \\
\frac {i}{2}(w_1-w_2)&=&r\sinh \theta \sin \varphi,
\nonumber \\
w_3&=&r\cosh \theta.
\end{eqnarray}
The infinitesimal generators corresponding to Eq. (1) in view of
Eqs. (3)-(5) read:
\begin{eqnarray}
{\cal K}_1&=&{\cal F}\partial_- + {\cal E}_-\partial_{{\cal F}},
\nonumber \\
{\cal K}_2&=&-{\cal F}({\cal E}_+\partial_++{\cal E}_-\partial_-)
+\frac {1}{2}(1+{\cal E}_+^2-{\cal E}_-^2-{\cal F}^2)\partial_{{\cal F}},
\nonumber \\
{\cal K}_3&=&{\cal E}_-({\cal E}_+\partial_++{\cal F}\partial_{{\cal F}})
+\frac {1}{2}(1+{\cal E}_+^2+{\cal E}_-^2+{\cal F}^2)\partial_-,
\nonumber \\
{\cal K}_0&=&{\cal E}_+({\cal F}\partial_{\cal F}+{\cal E}_-\partial_-)
+\frac {1}{2}(1+{\cal E}_+^2+{\cal E}_-^2-{\cal F}^2)\partial_+.
\nonumber
\end{eqnarray}
Here generator ${\cal K}_1$ corresponds to the parameter
${\frac {\lambda_1}{2}}$, etc.;
$\partial _{\pm}=\frac {\partial}{\partial _{{\cal E}_{\pm}}}$ and
we write down only the holomorphic parts of the generators. The calculation
of commutation relations gives:
\begin{eqnarray}
[{\cal K}_1, {\cal K}_2]=-{\cal K}_3, \quad
[{\cal K}_2, {\cal K}_3]={\cal K}_1, \quad
[{\cal K}_3, {\cal K}_1]={\cal K}_2,
\end{eqnarray}
and ${\cal K}_0$ commutes with ${\cal K}_1$, ${\cal K}_2$ and ${\cal K}_3$.

Next, from Eqs. (7), (10) and (13) it follows that
\begin{eqnarray}
\hat {\cal K}_1&=&\cos \varphi \partial_{\theta}-\coth \theta \sin \varphi
\partial_\varphi,
\nonumber \\
\hat {\cal K}_2&=&\sin \varphi \partial_{\theta}+\coth \theta \cos \varphi
\partial_\varphi,
\nonumber \\
\hat {\cal K}_3&=&-\partial_\varphi,
\nonumber \\
\hat {\cal K}_0&=&r\partial_r.
\nonumber
\end{eqnarray}
One can see that the group structure becomes evident in terms of the
variables $r$, $\theta$ and $\varphi$. The generators $\hat {\cal K}_1$,
$\hat {\cal K}_2$ and
$\hat {\cal K}_3$ depend only on the complex ``angles'' $\theta$ and
$\varphi$;
they define ``tangential'' transformations. Operator $\hat {\cal K}_0$
generates a ``transversal'' movement along the ``radial coordinate'' $r$.
From this geometric picture the commutation of
two subgroups immediately follows.

One can prove that the commutators of generators with hats are the
same as the commutators of generators without hats. Using this fact we
identify them correspondingly, i.e. we put
$\hat {\cal K}_1={\cal K}_1$, etc. We obtain an explicit form of
the relations (11) in the following way. First, we identify
$\hat {\cal K}_3$ with ${\cal K}_3$ (from the commutation relations
we see that the third vector is special). Second, we compare
$\hat {\cal K}_1$ and ${\cal K}_1$ (then the equality
$\hat {\cal K}_2={\cal K}_2$ becomes an identity
accordingly to Eqs. (14)). After the first two steps we will know the
dependence of ${\cal E}_{\pm}$ and ${\cal F}$ on $\theta$ and $\varphi$.
Finally, the comparison of the zero vectors gives the remaining
$r$-dependence.

Now let us briefly discuss details of the solution procedure. The equation
$\hat {\cal K}_3={\cal K}_3$ is equal to the system
\begin{eqnarray}
{\cal E}_{+,\varphi}&=&-{\cal E}_-{\cal E}_+,
\nonumber \\
{\cal F}_{,\varphi}&=&-{\cal E}_-{\cal F},
\nonumber \\
-2{\cal E}_{-,\varphi}&=&1+{\cal E}_+^2+{\cal E}_-^2+{\cal F}^2.
\end{eqnarray}
To solve this system it is useful to introduce the function
$\Psi = \sqrt {{\cal E}_+^2+{\cal F}^2}$. Then from Eqs. (15) one obtains
simple equations for ${\cal E}_-$ and $\Psi$. Solving them, it is easy to
calculate ${\cal E}_+$ and ${\cal F}$:
\begin{eqnarray}
{\cal E}_-&=&
-\frac {\sin (\varphi + \alpha)}{\cos \beta + \cos (\varphi + \alpha)},
\nonumber \\
{\cal E}_+&=&
\frac {\sin \beta \cos \gamma}{\cos \beta + \cos (\varphi + \alpha)},
\nonumber \\
{\cal F}&=&
\frac {\sin \beta \sin \gamma}{\cos \beta + \cos (\varphi + \alpha)},
\end{eqnarray}
where $\alpha$, $\beta$ and $\gamma$ are functions of $\theta$ and $r$.
Next, the relation $\hat {\cal K}_1={\cal K}_1$ leads to the following
extraction of the $\theta$-dependence:
\begin{eqnarray}
\alpha &=& \frac {\pi}{2},
\nonumber \\
\cos \beta &=& i\frac {\cos \Lambda}{\sinh \theta},
\nonumber \\
\sin \beta \cos \gamma &=& i\frac {\sin \Lambda}{\sinh \theta},
\nonumber \\
\sin \beta \cos \gamma &=& \coth \theta,
\end{eqnarray}
where $\Lambda$ is an arbitrary function of $r$.

The last step consists of a comparison of the zero vectors. However, zero
vectors commute with the other ones, and commutation relations only suggest
that
\begin{eqnarray}
\hat {\cal K}_0=C{\cal K}_0,
\end{eqnarray}
where $C$ is a constant. A straightforward calculation shows that the
identification of the zero vectors leads to the relation $\Lambda = C\ln r +
D$.
Now we choose the values of $C$ and $D$ to provide the simplest (rational)
form of
the functions (11) and to make the trivial point $w_1=w_2=w_3=0$
corresponded to the vacuum configuration. The result is:
\begin{eqnarray}
C=-i, \quad D=0.
\end{eqnarray}

Finally, from Eqs. (12), (13) and (16)-(19) one obtains the explicit
form of the
relations between MEP components and linearizing potentials:
\begin{eqnarray}
{\cal E}&=&
i\frac {(1-w_1)(1-w_2)-w_3^2}{(1+w_1)(1-w_2)+w_3^2},
\nonumber \\
{\cal Z}&=&
i\frac {(1+w_1)(1+w_2)-w_3^2}{(1+w_1)(1-w_2)+w_3^2},
\nonumber \\
{\cal F}&=&
-\frac {2iw_3}{(1+w_1)(1-w_2)+w_3^2}.
\nonumber
\end{eqnarray}
The inverse relations have the form:
\begin{eqnarray}
w_1&=&
\frac {1+{\cal E}{\cal Z}+{\cal F}^2+i({\cal E}-{\cal Z})}
{1-{\cal E}{\cal Z}-{\cal F}^2-i({\cal E}+{\cal Z})},
\nonumber \\
w_2&=&-
\frac {1+{\cal E}{\cal Z}+{\cal F}^2-i({\cal E}-{\cal Z})}
{1-{\cal E}{\cal Z}-{\cal F}^2-i({\cal E}+{\cal Z})},
\nonumber \\
w_3&=&
\frac {2i{\cal F}}
{1-{\cal E}{\cal Z}-{\cal F}^2-i({\cal E}+{\cal Z})},
\end{eqnarray}

The new potentials $w_1, w_2$ and $w_3$ seem to be the most suitable for the
study of charged stationary EMDA configurations possessing
asymptotic flatness property. Actually, they undergo simple transformation 
rules and also
\begin{eqnarray}
\hat {\cal W}_{as}=\frac {\hat {\cal M}}{r},
\nonumber
\end{eqnarray}
i.e. these potentials are free of any non-vanishing asymptotics (from
this relation we see how the formulae (6) and (10) can have      identical
form).
\section{Concluding remarks}
Thus, the stationary EMDA allows a representation which linearizes the
action of its charging symmetry subgroup. The potentials that
realize this representation do not form the
analogy of Kinnersley's linearizing potentials
\cite {kin}. Actually, to linearize the $complete$ $symmetry$ $group$ of the
stationary EM theory Kinnersley introduced a new potential in
addition to the present pair of potentials. In our case, the result is
different: we linearize only the $charging$ $symmetry$ $subgroup$ without 
extension of the potential space and use only the appropriate change of 
variables.

The close relation between linearizing potentials and charges allow us to
establish one general invariant of the charging symmetry subgroup.
Actually, the charge function
$I({\cal M}, {\cal D}, {\cal Q})=|{\cal M}|^2
+|{\cal D}|^2-2|{\cal Q}|^2 \equiv {\rm Tr}\left ( \bar {\hat {\cal M}}
\sigma_3 \hat {\cal M} \sigma_3\right )=I(\hat {\cal M})$
is invariant
under the action of charging symmetries in view of Eqs. (6) and (9) (this
invariant vanishes for BPS-saturated configurations). 
Identical transformation properties of $\hat {\cal M}$ and $\hat {\cal W}$
allow us to introduce the corresponding invariant function of the
linearizing potentials
$I(\hat {\cal W})=
{\rm Tr}\left ( \bar {\hat {\cal W}}\sigma_3 \hat {\cal W} \sigma_3\right )
=|w_1|^2+|w_2|^2-2|w_3|^2=I(w_1, w_2, w_3)$. From Eq. (20) it follows
that in terms of the EMDA Ernst potentials the charging symmetry invariant 
takes the form
\begin{eqnarray}
\frac {1}{2}I({\cal E}, {\cal Z}, {\cal F})=
\frac {|1+{\cal E}{\cal Z}+{\cal F}^2|^2+
|{\cal E}-{\cal Z}|^2-4|{\cal F}|^2}
{|1-{\cal E}{\cal Z}-{\cal F}^2-i({\cal E}+{\cal Z})|^2}.
\nonumber
\end{eqnarray}
One can see that the first two terms of the $1/r$ expansion of this invariant
vanish for the BPS-saturated fields.
\section*{Acknowledgments}
We would like to thank our colleagues for encouraging us in our work.
One of the authors (A.H.) was partially supported by CONACYT and SEP.

\end{document}